\documentclass[bibyear, letter]{aa}
\usepackage{graphicx}
\usepackage{txfonts}
\usepackage[colorlinks=true, 
urlcolor=blue, 
citecolor=blue, 
linkcolor=blue]{hyperref}
\usepackage{natbib}
\usepackage{booktabs}
\usepackage{xcolor}
\usepackage[normalem]{ulem} 
\usepackage[separate-uncertainty=true]{siunitx} 
\usepackage{float}
\usepackage{orcidlink}
\usepackage[switch]{lineno}
\begin{document}

   \title{Early emission characterization of TDE~2025aarm}

   \titlerunning{Early-time characterization of TDE~2025aarm}

   \author{
    Andrea Simongini$^{\orcidlink{0009-0000-3416-9865}}$,$^{1,2}$\thanks{E-mail: andrea.simongini@inaf.it}
    Maria Kherlakian$^{\orcidlink{0000-0003-4686-0922}}$,$^{3}$ 
    Alicia L\'opez-Oramas$^{\orcidlink{0000-0003-4603-1884}}$, $^{4}$
    Josefa Becerra$^{\orcidlink{0000-0002-6729-9022}}$ $^{4, 5}$ 
    \and 
    Davide Cerasole$^{\orcidlink{0000-0003-2033-756X}}$ $^{6, 7}$ 
    }
    \institute{
    $^{1}$ INAF - Osservatorio Astronomico di Roma, Via di Frascati 33, I-00078 Monteporzio Catone, Italy\\
    $^{2}$ Università Tor Vergata, Dipartimento di Fisica, Via della Ricerca Scientifica 1, I-00133 Rome, Italy \\
    $^{3}$ Astronomical Institute, Faculty for Physics and Astronomy, Ruhr University Bochum, Bochum, 44780, Germany \\ 
    $^{4}$ Instituto de Astrofísica de Canarias and Departamento de Astrofísica, Universidad de La Laguna, C. Vía Láctea, s/n, 38205 La Laguna, Santa Cruz de Tenerife, Spain\\
    $^{5}$ Consejo Superior de Investigaciones Científicas (CSIC), E-28006 Madrid, Spain\\ 
    $^{6}$ Dipartimento di Fisica ``M. Merlin" dell'Universit\`a e del Politecnico di Bari, I-70126 Bari, Italy \\
    $^{7}$ Istituto Nazionale di Fisica Nucleare, Sezione di Bari, I-70126 Bari, Italy
    }
    \authorrunning{A. Simongini et al.}

    \date{Received 2 March 2026; accepted 2 June 2026}

\abstract{

    In this work, we present early emission data analysis of the tidal disruption event TDE~2025aarm, including optical, UV, and X-ray data. 
    At a redshift of $z = 0.01368$, TDE~2025aarm is the second closest TDE ever discovered, offering a valuable opportunity to study such phenomena in great detail. 
    We observed TDE~2025aarm in the optical with the Liverpool Telescope for a total of three epochs. We complemented our dataset with ancillary spectroscopic and photometric data. 
    The early optical spectra are characterized by a blue-continuum and helium, hydrogen, and possibly Bowen lines typical of H+He events. 
    The optical light curves peak at $M_g \sim -18.68$ mag and are well described by fallback of a $M_\star\sim0.16 M_\odot$ star onto a $M_{\rm BH} \sim 2\times10^{7}M_\odot$ black hole.     
    We report \textit{Swift}-XRT detection in the $0.3-10$ keV range, with a total flux of $F_{X} \sim 1.42\times10^{-14}\rm\,erg \, cm^{-2}\,s^{-1}$, fit by a blackbody with $k_BT \sim 0.39$ keV. 
    This makes TDE~2025aarm a new event among optical/UV bright TDEs detected in soft X-rays. 
    Our analysis suggests that the early emission from TDE~2025aarm is powered by circularization shocks and that the delayed accretion scenario best describes the observed features.

}


    \maketitle

\nolinenumbers

%-------------------------------------------------------------------

\section{Introduction}

    Tidal disruption events (TDEs) are transient phenomena that occur when stars wander sufficiently close to a supermassive black hole (BH) to be torn apart by tidal forces.
    A luminous flare of radiation, with typical luminosities of $10^{42} - 10^{44}\rm\,erg\,s^{-1}$, is expected from the fraction of stellar debris that, being gravitationally bound after disruption, fall back onto the BH and form an accretion disk~\citep{hills1975possible, rees1988tidal}. 
    Over the past decade, wide-field optical and X-ray surveys have discovered more than 100 TDEs, establishing them as a distinct class of luminous nuclear transients (e.g., \citealt{hammerstein2023final}). 
    These events typically exhibit blue optical/UV (OUV) continua and light curves that rise over days to weeks and decline over weeks to months, which are often approximated by a power-law decay consistent with the fallback rate of stellar debris onto the accretion disk (e.g., \citealt{nicholl2022systematic}). 
    Spectroscopically, TDEs are divided into four main classes, He, H+He, H, and featureless (e.g., \citealt{charalampopoulos2022detailed, hammerstein2023final}), with one additional less-emerging class discovered in recent years: TDE coronal \citep{yao2023tidal}. 
    The TDEs in the H+He class tend to prefer more OUV luminous flares with more compact radii than TDE H \citep{van2021seventeen}. The TDEs in the He class emerge for sufficiently high luminosities and compact radii to ionize \ion{He}{ii}, while featureless spectra are expected for the highest luminosities \citep{aspegren2026emission}. 
        
    Though TDEs are expected to be mainly powered by accretion, almost 50$\%$ of the discovered TDEs exhibit faint X-ray radiation or none at all. 
    There are two main models that attempt to explain this diversity. 
    In the first scenario, the accretion disk formation is delayed, while the OUV emission is powered by shocks from debris stream–stream collisions near apocenter during the circularization process. The dissipated kinetic energy is thermalized and diffuses through the infalling debris, producing bright OUV emission that is largely decoupled from the X-ray emission from the inner accretion flow \citep{piran2015disk}. 
    In the second scenario, the circularizing debris form a viscous accretion disk around the supermassive BH (SMBH), which can effectively obscure the X-ray emission and reprocess it into OUV wavelengths \citep{rees1988tidal}, with viewing angle effects potentially intervening \citep{dai2018unified}.

    The paper is structured as follows: we present the discovery of TDE~2025aarm and the status of multi-wavelength observations in Sect.~\ref{sec:discovery}. 
    In Sect.~\ref{Sec:2} we present the optical, UV, and X-rays data used in this analysis. 
    In Sect.~\ref{Sec:analysis} we focus on analyzing the spectral features, reconstructing the photospheric parameters, and modeling the light curves of the event. 
    Finally, in Sect.~\ref{Sec:7} we discuss our results and draw the final conclusions.

\section{Discovery}\label{sec:discovery}

    TDE~2025aarm (AT 2025aarm) was discovered by GOTO on MJD 60949.17 with a L-filter magnitude of 18.96 at the coordinates RA, DEC (J2000) = $68.05, -5.38$ deg \citep{ONeill2025}. 
    It is located 0.186'' from the center of its host galaxy (LEDA 3681212; \citealt{fabricius2021gaia}), indicating that the event is coincident with the galaxy center within subarcsecond precision.
    It was classified as a TDE H+He at a redshift of $z = 0.01368$ \citep{Newsome2025}. 
    Assuming a flat $\Lambda$CDM Universe with $H_0 = 67.4\,\rm km\,s^{-1}\,Mpc^{-1}$, $\Omega_m = 0.315$, and $\Omega_\Lambda = 0.685$ \citep{2020A&A...641A...6P}, we obtained a luminosity distance of $D_{\rm L} = 61.48 \pm 6.15$ Mpc, making TDE~2025aarm the second closest TDE ever discovered, with only AT~2023clx \citep{charalampopoulos2024fast} being closer.
    The distance modulus is $\mu = 33.94 \pm 0.22$ mag. 
    
    Owing to its vicinity, TDE~2025aarm was and still is monitored across the entire electromagnetic spectrum. 
    A first non-detection in the 1 -- 8 GHz radio bands was reported by \citet{Sfaradi2025} with the Allen Telescope. 
    It followed a significant detection with the Very Large Array at 15.1 GHz with a flux density of $36\,\rm\mu Jy$, corresponding to a luminosity of $L_{\rm 15\,GHz} =  1.6 \times 10^{26}\rm\,erg\,s^{-1}\,Hz^{-1}$ and a positive spectral index \citep{Christy2025}. 
    The source was significantly detected in the 0.5 -- 7.0 keV X-ray band with the Chandra Telescope, with an unabsorbed flux of $5.9~\times~ 10^{-15}\rm\,erg\,cm^{-2}\,s^{-1}$, a spectral index of 2.2, and a luminosity of $L_{\rm 0.5 - 7\,keV} = 2.5\times10^{39}\rm\,erg\,s^{-1}$ \citep{Somalwar2025}. 
    Observations were also performed in the GeV--TeV gamma-ray bands, though without achieving any significant detection \citep{Mohrmann2025, Paneque2025}. 
    
\section{Observations and data}\label{Sec:2}

\subsection{Optical}

    We observed the position of TDE~2025aarm for three epochs (MJD 61016.93, 61030.04, 61084.90) with the SPectrograph for the Rapid Acquisition of Transients (SPRAT; \citealt{piascik2014sprat}) at the Liverpool Telescope (LT) via Director's Discretionary Time (IDs CQ25B01 -- CQ26A01; PI: A. L\'opez-Oramas). 
    Observations were performed with a blue-optimized setup, and we used the standard LT automatic pipeline for data reduction. 
    See Table~\ref{tab:spectra} for details. 
    We collected one additional public spectrum from \texttt{WISeREP} \citep{Yaron2017}. 
    This spectrum was observed on MJD 60977.43 with the FLOYDS-N at the Las Cumbres Observatory (LCO; PI: M. Newsome). 

    \begin{figure}
    \centering
    \includegraphics[width=\columnwidth]{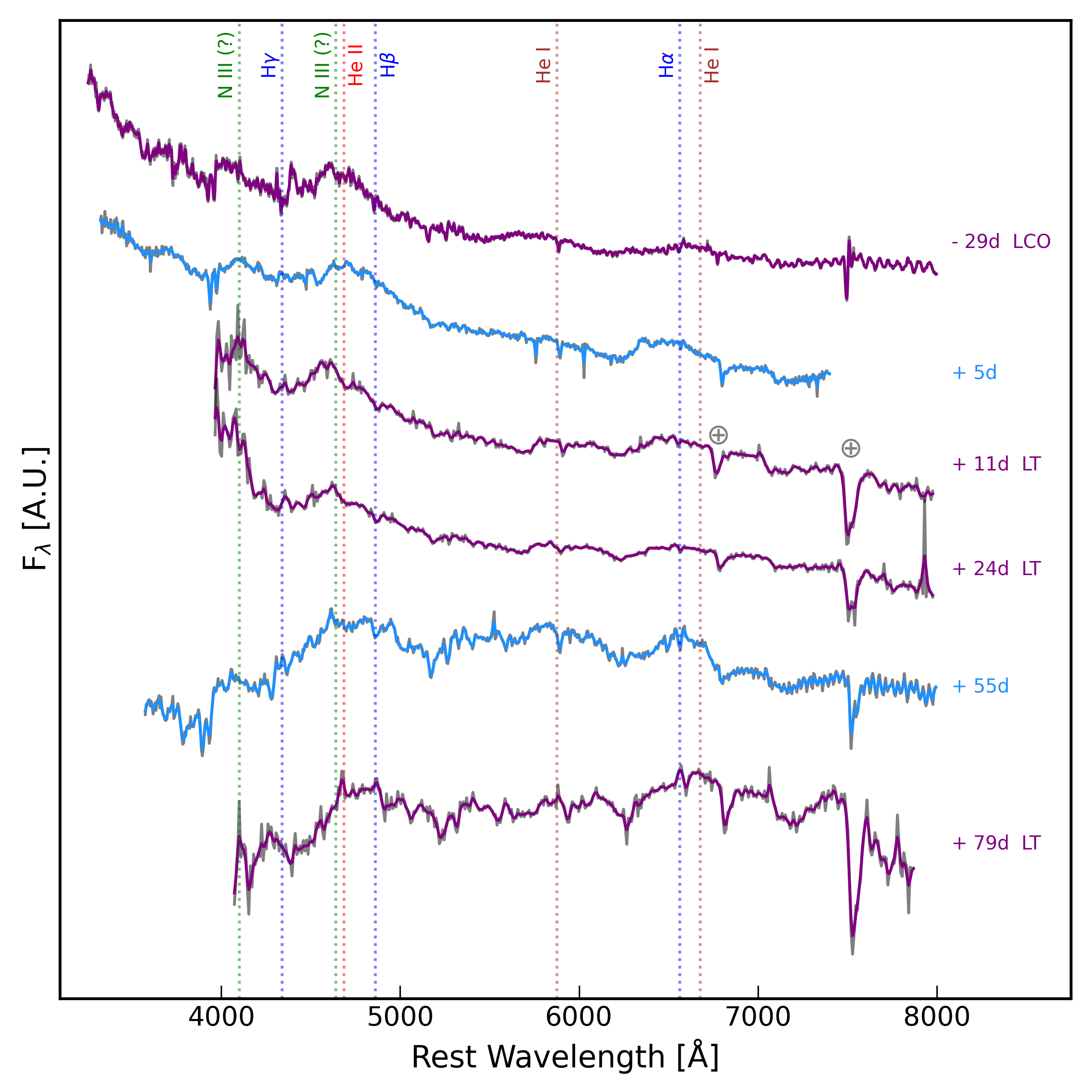}
    \caption{Spectral evolution of TDE~2025aarm between -29 and +79 days from $t_{\rm p}$. 
    We show the original spectra in gray and the smooth interpolated in purple.
    For comparison, we show two spectra of AT~2023clx in blue. 
    Fluxes have been normalized and offset for visual clarity. 
    The dashed vertical lines are placed on the rest-frame position of the relevant spectral features.}
    \label{fig:spectra}
    \end{figure}

    We collected ancillary photometric data from public brokers, including $g$ and $r$ filter forced-photometry from the Zwicky Transient Facility \citep{graham2019zwicky} and $o$ and $c$ filter photometry from ATLAS \citep{tonry2018atlas}. 
    The light curves cover a total range between MJD 60939.42 and 61086.78. 
    The complete photometry is reported in Table~\ref{tab:uvot} and shown in Fig.~\ref{fig:lightcurves}. 
    Every magnitude is expressed in the AB system and has been converted into absolute magnitudes after cosmological K correction $2.5\,\rm log_{10}(1+z)$ and considering a Milky Way extinction of $E_{B-V} = 0.0688$ \citep{schlafly2011measuring}. 
    We derived the values at the peak by smoothly interpolating the data points, obtaining $M_{\rm peak, g} = -18.68 \pm 0.03\,\rm mag$ at $t_{\rm peak, g} = 61006.38 \pm 3.26\,\rm MJD$ and $M_{\rm peak, r} = -18.41 \pm 0.2\,\rm mag$ at $t_{\rm peak, r} = 61010.44 \pm 2.95\,\rm MJD$.
    We use $t_p = t_{\rm peak, g}$ as the reference epoch throughout this work. 

    \subsection{\textit{Swift}-UVOT}

    Target of opportunity observations spanning several epochs between MJD 60979 and 61078 (PIs: Kuin, Stein, Miller, Charalampopoulos, Sun, Konno) were obtained with the UV-Optical Telescope (UVOT; \citealt{roming2005swift}) and X-ray Telescope (XRT; \citealt{burrows2005swift}) on board the Neil Gehrels Swift Observatory.
    We reduced the UVOT photometry via the \texttt{uvotredux} \citep{robert_david_stein_2025_17634569} \texttt{python} wrapper, which is based on the \texttt{uvotsource} module of the \texttt{HEASoft} package \citep{2014ascl.soft08004N}.
    We subtracted the host contribution modeled by the \texttt{prospector} package~\citep{prospector_package}, based on publicly available data of the host galaxy (see Appendix~\ref{sec:prospector}). 
    After subtraction, the brightest filter is $w_2$, reaching a peak of $M_{\rm peak, w_2} = -19.32 \pm 0.02\,\rm mag$ at $t_{\rm peak, w_2} = 60988.42 \pm 3.53\,\rm MJD$

\subsection{\textit{Swift}-XRT}

    We searched for X-ray emission from the position of TDE~2025aarm in the XRT data. 
    Notably, only upper limits were reported from the host galaxy \citep{2025ATel17466....1M}. 
    We report a stacked detection with a count rate of $(4.3\pm1.2) \times 10^{-4}\rm\,cts\, s^{-1}$ for a total exposure time of 59.2 ks between MJD 60979 and 61078.  
    We used the \texttt{Xpsec} module \citep{arnaud1996astronomical} within \texttt{HEASoft} to fit the spectrum, assuming a Galactic column density of $4.4 \times 10^{20}\rm\, cm^{-2}$ \citep{bekhti2016hi4pi}. 
    Both power-law and blackbody models fit the data with reduced statistics of 0.87 and 0.93, respectively (see Appendix~\ref{sec:xrt}). 
    The resulting unabsorbed flux in the $0.3-10$ keV range for the blackbody model is $F_{0.3-10\rm\, keV} = (1.42\pm0.36) \times 10^{-14}\,\rm erg\,cm^{-2}\,s^{-1}$, with a temperature of $k_BT = 0.39\pm0.08$ keV, corresponding to a luminosity of $L_{0.3-10\rm\, keV} = (6.42\pm2.07) \times 10^{39}\,\rm erg\,\,s^{-1}$. 
    The power-law model yields a photon index of $\Gamma = 1.96\pm0.4$, consistent with Chandra \citep{Somalwar2025}. 
    
    Notably, by dynamically binning the counts across the entire time range, we obtained two bins above the threshold at 61004 MJD and 61065 MJD, with the second at a higher level, suggesting that the flux is increasing.
    Consistent with this trend, a second detection by Chandra on 61141 MJD confirmed the X-ray brightening by a factor of about ten compared to the XRT flux \citep{2026TNSAN.101....1J}. 
    The measured unabsorbed flux in the 0.5 -- 7 keV band is $1.4\pm2\times10^{-13}\,\rm erg \, cm^{-2}\,s^{-1}$ for a power-law model with photon index $1.6\pm0.4$, corresponding to a luminosity of $6.3\times10^{40}\rm\,erg\,s^{-1}$.

\section{Analysis}\label{Sec:analysis}

\subsection{Spectral analysis}

    We identified the spectral emission features of TDE~2025aarm by comparing our spectra with those of AT~2023clx \citep{charalampopoulos2024fast}. 
    In both cases, the spectra exhibit a blue-continuum on top of broad emission lines, which cools down tens of days after the peak (Fig.~\ref{fig:spectra}). 
    Both sources display typical H+He class emission lines, including the Balmer series and the helium ions \ion{He}{i} $\lambda\lambda 5876, 6678$ and \ion{He}{ii} $\lambda 4684$. 
    The presence of Bowen fluorescence features cannot be ruled out nor robustly confirmed. 
    The \ion{N}{iii} $\lambda 4640$ is strongly blended with \ion{He}{iii}, while \ion{N}{iii} $\lambda 4100$ is potentially blended with \ion{H$\delta$}{} (see Fig.~\ref{fig:spectral_fit}). 

    We subtracted the host light using the spectrum from DESI Data Release 1 (ID: 2842392105320449; \citealt{abdul2025data}). 
    We fit the spectral features of the continuum-subtracted spectra with \texttt{PyQSOFit} \citep{2018ascl.soft09008G} to extract their relevant information, luminosity, full width at half maximum (FWHM), and offset as well as to track their evolution (Fig.~\ref{fig:lines}).  
    The behavior of the lines after maximum is in concert with the described properties of TDEs H+He.
    They are broad, with an FWHM on the order of $10^3 - 10^4 \rm\, km \, s^{-1}$ and an increasing broadening over time. 
    The luminosity of each line increases with time, reaching the peak between 11 -- 24 days post $t_{\rm p}$, although sparse sampling hinders the exact estimation of the time lag. 
    The lag time roughly follows $\tau \sim r/c$ \citep{charalampopoulos2022detailed}. 
    For TDE~2025aarm this translates into $r = 7 \times 10^{16}$ cm, which is two orders of magnitude higher than the blackbody radius. 
    \cite{charalampopoulos2022detailed} noticed that the evolution of the spectral lines and the photospheric evolution are correlated, and the lag can be interpreted as low electron density reprocessing material with high recombination times.
    The correlation between spectral lines and the photosphere can also be seen in the \ion{He}{ii}/H$\alpha$ ratio, which is higher when the radius is expanding (receding) before (after) the peak and lower when the radius is at its peak. 
    This hints at a stratified emitting structure, with Helium closer to the BH that becomes stronger when the photosphere recedes. 
    Finally, all the lines are red- or blue-shifted and have a generally decreasing trend after the peak toward null.

\subsection{Blackbody fit}

    We studied the evolution of the optical photon field using \texttt{extrabol} \citep{thornton2024extrabol}, which reconstructs the bolometric luminosity, the blackbody temperature, and radius (Fig.~\ref{fig:photosphere}). 
    See Appendix~\ref{sec:extrabol} for more details. 
    
    The bolometric luminosity peaks at -11 days, reaching $L_{\rm bb, \,peak} = (3.4\pm0.1)\times10^{43}\,\rm erg \, s^{-1}$, corresponding to a peak bolometric magnitude of $M_{\rm peak} = -20.13\pm0.03$ mag.  
    The temperature exhibits a first peak of $T_{\rm bb, \, peak_1} = (22.16\pm0.3)\times10^3\,\rm K$ at -24 days and then drops quickly after. 
    It rises again at 37 days, reaching $T_{\rm bb, \, peak_2} = (20.33\pm 1.0)\times10^3\,\rm K$, corresponding to a second bump in the luminosity curve. 
    The photospheric radius expands during the first $\sim$ 60 days after discovery, reaching a maximum extension of $R_{\rm bb, \, peak} = (6.6\pm0.1)\times10^{14}\,\rm cm$, and it exhibits similar bumps as the temperature and luminosity curves. 
    An equivalent behavior has been observed in other TDEs, such as ASASSN-15oi \citep{gezari2017x}, ASASSN-15lh \citep{leloudas2016superluminous}, and AT~2018fyk \citep{wevers2019evidence}.

\subsection{Fallback modeling}

    After tidal disruption, a fraction of the stellar mass is gravitationally bound to the BH. 
    The subsequent fallback onto the BH can be subdivided into two phases: circularization and accretion. 
    The resulting radiation directly depend on the BH mass, $M_{\rm BH}$, by
    \begin{equation}
        \frac{dM}{dE}\frac{dE}{dt} = \frac{2\pi}{3} \left(G M_{\rm BH}\right)^{2/3}\frac{dM}{dE} t^{-5/3}.
    \end{equation}
    This equation describes a complete fallback. If the fallback is partial, we expect a steeper rate of $t^{-9/4}$. 
    Depending on the mass of the BH, the accretion rate can be in the super-Eddington or in the sub-Eddington regime. 
    In particular, the relation that describes the Eddington luminosity is 
    \begin{equation}
        L_{\rm Edd} = \eta \dot M_{\rm Edd} c^2 = 1.33 \times 10^{44} \left(\frac{M_{\rm BH}}{10^6 M_\odot}\right)\ \rm erg\, s^{-1}
    ,\end{equation}
    where $\eta$ is the accretion efficiency. 
    See, for example, \cite{strubbe2009optical} for additional details on this model. 

    We modeled our light curves using the \texttt{tde} module from \texttt{MOSFiT} \citep{guillochon2018mosfit}. 
    We show the results in Fig.~\ref{fig:mosfit} and report the prior and posterior of the free parameters in Table~\ref{table:mosfit}. 
    The model reproduces well most of the observed data, with the lowest chi-square yielded for the $g$-band $\chi^2_g$/d.o.f. = 0.22 and the $uvm2$-band $\chi^2_{w1}$/d.o.f. = 0.99. 
    The $U$-band yields the highest residuals, and in general, the UVOT filters are the hardest to reproduce, as seen in \citet{nicholl2022systematic}. 
    The fit yields a total reduced chi-square of $\chi^2$/d.o.f. = 1.54. 
    Our results suggest a low-mass star, $M_\star = 0.16^{+0.03}_{-0.03}\, M_\odot$, disrupted by a high-mass BH, $M_{\rm BH} =  1.87^{+0.3}_{-0.2}\times10^{7} M_\odot$, with a scaled impact factor of $b \sim 0.5$ and a relatively low efficiency of $\epsilon \sim 0.05$, which is in line with the population of H+He TDEs studied in \citet{nicholl2022systematic}. 
    These results indicate sub-Eddington accretion, with $L_{\rm OUV}/L_{\rm Edd} = 0.01$ and $L_{X}/L_{\rm Edd} = 3\times10^{-6}$,  and reveal an extremely underluminous X-ray emission. 
    However, since \texttt{MOSFiT} measurements show only a marginal agreement with BH masses inferred from host-galaxy properties (see, e.g., \citealt{guolo2025compact}), these values are not to be taken as exact estimates. 
    In this regard, the independent measurement by \texttt{prospector} (see Sect.~\ref{sec:prospector}) provides additional proof of the order of magnitude of the SMBH behind TDE~2025aarm.

\section{Discussion and conclusions}\label{Sec:7}

    TDE~2025aarm adds to the diversity of OUV bright TDEs that have been detected in the soft X-ray band \citep{van2020optical, hammerstein2023final}. 
    Remarkably, TDE~2025aarm exhibits one of the highest OUV to X-ray luminosity ratios at peak ever found \citep{hammerstein2023final}, with $L_{\rm OUV}/L_X = 5.3\times10^{3}$. 
    The faint X-ray emission observed in TDE~2025aarm may reflect inefficient accretion from a still-circularizing debris stream.   
    This behavior was potentially already hinted at by the increasing counts in the XRT analysis, although the statistics are still too low to provide solid evidence.
    Alternatively, the X-ray emission could be obscured by optically thick material, although the early detection by Chandra may suggest only partial veiling, as discussed for other TDEs \citep{gezari2017x}. 
    
    Interestingly, the delayed accretion scenario is potentially probed by the bumps in the temperature evolution (Fig.~\ref{fig:photosphere}). 
    Here, the first peak reflects radiation coming from circularization shocks, while the subsequent re-brightening is due to the prompt and efficient formation of the accretion disk and consequently an increase in reprocessing X-rays to OUV light \citep{leloudas2016superluminous, wevers2019evidence, charalampopoulos2024fast}.
    For TDEs occurring in high-mass SMBHs, such as TDE~2025aarm, the disk formation is expected to be more efficient and to happen within tens to hundreds of days after disruption \citep{wong2022revisiting}. 
    
    The spectral line evolution provides an additional argument. 
    For Bowen lines to emerge, there must be a source of X-rays. 
    In the orientation-dependent model,  where the reprocessing happens around the accretion disk, the line width and the blueshift positively correlate with X-ray emission and should rise as the photosphere becomes transparent. 
    This is the opposite of what we observed with TDE~2025aarm: The blueshift diminishes while the X-ray flux increases, indicating that the orientation scenario is not (the only) contributing factor. 
    In the delayed accretion scenario, instead, the blueshift is unrelated to changes in the X-ray flux. 
    Future X-ray observations of TDE~2025aarm will be crucial to testing these interpretations, as the delayed emergence of a luminous soft X-ray component is expected once a circularized accretion disk forms and viscous accretion becomes efficient or when the accreting material becomes optically thin (e.g., \citealt{gezari2017x, hammerstein2023final}). 
    Early signs that potentially support this interpretation are already in place \citep{2026TNSAN.101....1J}. 

    In conclusion, we have presented an early characterization of TDE~2025aarm. 
    Its proximity has enabled an exceptionally deep multiwavelength characterization, including the detection of soft X-ray emission that would likely have remained undetected at greater distances, 
    The high-cadence optical photometric and spectroscopic monitoring provides an unusually detailed view of its temporal and spectral evolution. 
    Future analyses will provide a more complete characterization of TDE~2025aarm. 
    For now, this work has established an initial observational framework that can serve as a reference for future studies, with particular emphasis on the importance of coordinated multi-wavelength follow-up observations.
    \\ 
    
    \noindent\small{\textit{Data availability} Table E.3 is only available in electronic form at the CDS via anonymous ftp to \url{cdsarc.u-strasbg.fr (130.79.128.5)} or via \url{http://cdsweb.u-strasbg.fr/cgi-bin/qcat?J/A+A/.}}
    
%--------------------------------------------------------------------
\begin{acknowledgements}

    The authors thank Patrik Veres for helpful discussion. 
    The Liverpool Telescope is operated on the island of La Palma by Liverpool John Moores University in the Spanish Observatorio del Roque de los Muchachos of the Instituto de Astrofisica de Canarias with financial support from the UK Science and Technology Facilities Council. 
    This research has made use of data and/or software provided by the High Energy Astrophysics Science Archive Research Center (HEASARC), which is a service of the Astrophysics Science Division at NASA/GSFC.
    ALO and JBG acknowledge support from the Agencia Estatal de Investigación del Ministerio de Ciencia, Innovación y Universidades (MCIU/AEI) under grant PARTICIPACIÓN DEL IAC EN EL EXPERIMENTO AMS and the European Regional Development Fund (ERDF) with reference PID2022-137810NB-C22. This work is part of the Project RYC2021-032991-I, funded by MICIU/AEI/10.13039/501100011033, and the European Union “NextGenerationEU”/PRTR.
\end{acknowledgements}

\bibliographystyle{aa}
\bibliography{thebib}

%--------------------------------------------------------------------
\begin{appendix}

\normalsize
\section{XRT analysis}\label{sec:xrt}

\begin{figure}[H]
    \centering
    \includegraphics[width=\columnwidth]{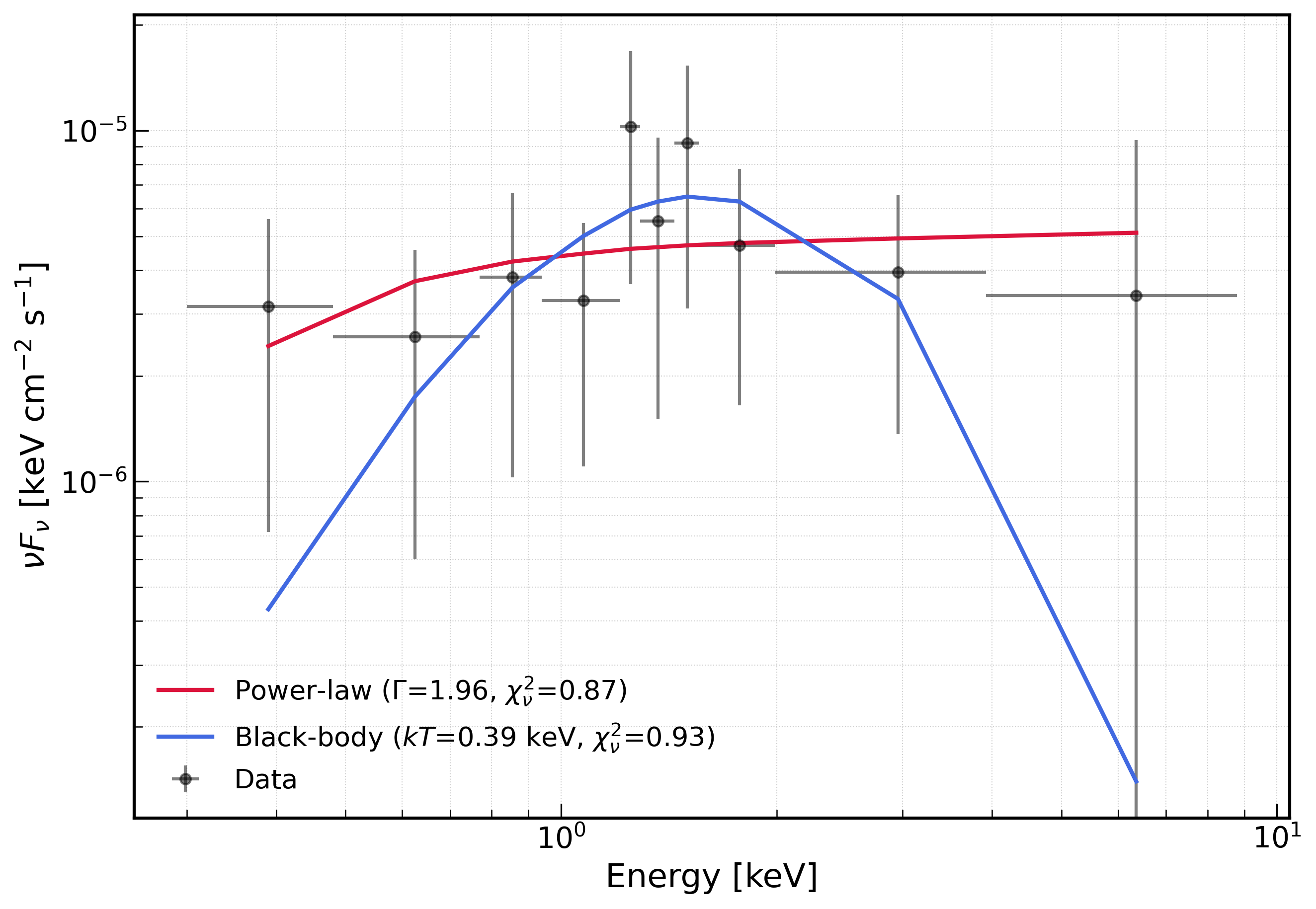}
    \caption{Spectral energy distribution fitting of the XRT data. Results for the power-law and blackbody models are shown in red and blue lines, respectively.}
    \label{fig:xrt_fit}
\end{figure}

The intrinsic X-ray spectral energy distribution (SED) of TDE~2025aarm reconstructed from the \textit{Swift}-XRT data is shown in Fig.~\ref{fig:xrt_fit}, where the correction of the spectral points for the Galactic extinction effects was performed as in \cite{abe2025very}. 
We fit both a blackbody and a power-law model, accounting for photon absorption and the nominal Galactic hydrogen column density of $4.4\times10^{20}\rm\, cm^{-3}$ \citep{bekhti2016hi4pi}. 
The fit was performed by minimizing the Cash statistic. 
Results of the fit with the associated statistics are reported in Table~\ref{tab:xrt}. 

\begin{table}[h!]
    \centering
    \caption{XRT fit results.}
    \label{tab:xrt}
    \begin{tabular}{l|cc}
    \toprule 
    \toprule 
         &  Power-law   & Blackbody \\ 
    \midrule 
    C-stat            &    38.45    &  40.89 \\ 
    Chi-squared       &    27.86    &  23.29 \\ 
    Reduced C-stat$^a$    &    0.87    &  0.93\\ 
    \midrule  
    Flux$^b$              &    2.47 $\pm$ 0.76 & 1.32 $\pm$ 0.34 \\ 
    Flux$^b$   (unabs)    &    2.74 $\pm$ 0.85 & 1.42 $\pm$ 0.36 \\
    Luminosity$^c$        &    12.4 $\pm$ 4.57 & 6.42 $\pm$ 2.07 \\ 
    \bottomrule 
    \end{tabular}
    \tablefoot{$^{a}$ With 44 degrees of freedom. $^{b}$Fluxes are expressed in units of $10^{-14}\rm\, erg \, cm^{-2}\, s^{-1}$. $^{c}$ Luminosities are estimated from the unabsorbed flux values, in units of $10^{39}\rm\, erg \, s^{-1}$.}
\end{table}

The resulting power-law has a photon index of $\Gamma = 1.96 \pm 0.4$, while the blackbody temperature is $k_{\rm B} T = 0.39 \pm 0.08$ keV, where the uncertainties are estimated with 200 simulations and a 68$\%$ confidence level.

\section{\texttt{extrabol} fitting}\label{sec:extrabol}

\begin{figure}[H]
    \centering
    \includegraphics[width=\columnwidth]{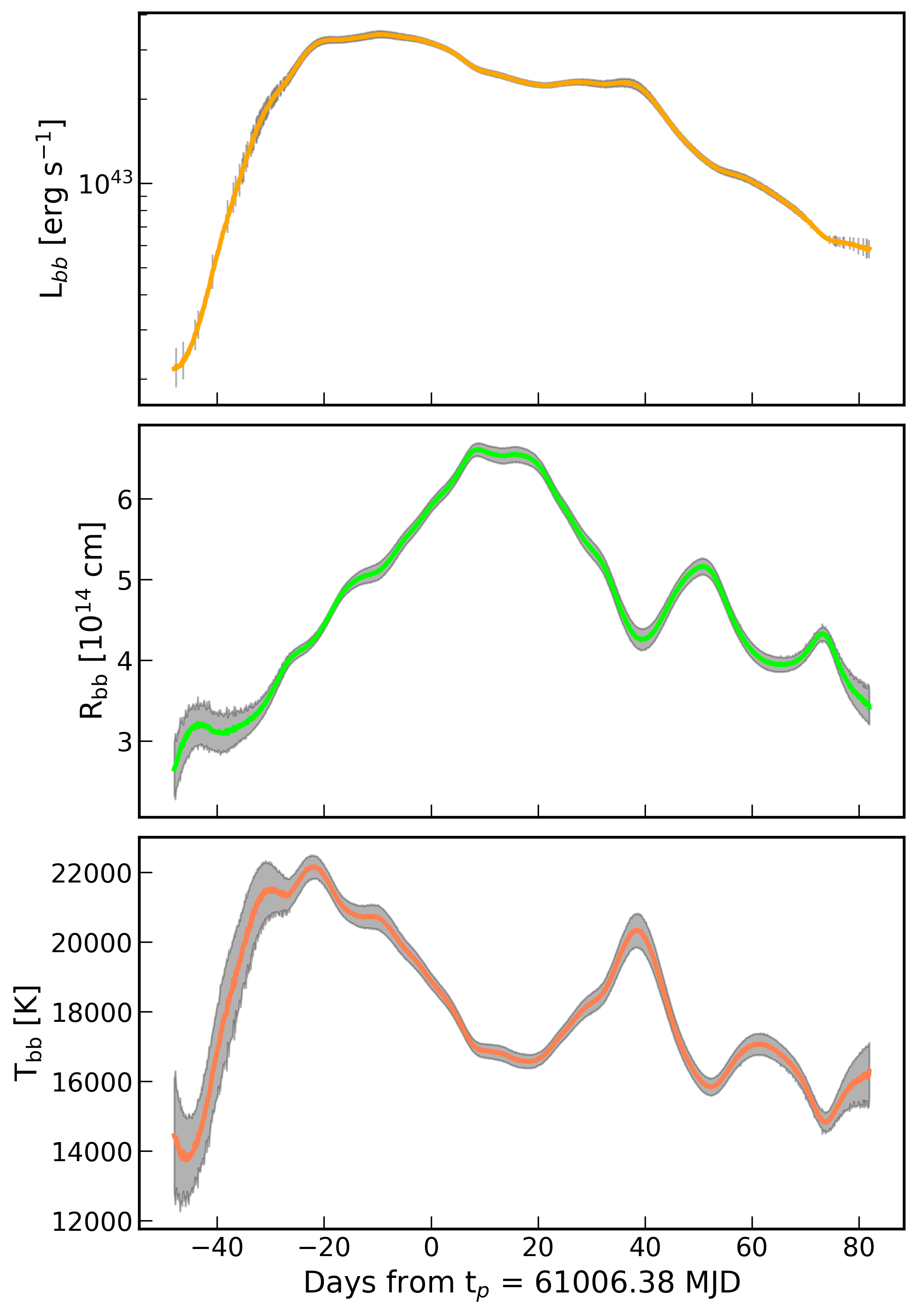}
    \caption{Photospheric evolution of TDE~2025aarm. \textit{Top}: Bolometric luminosity. \textit{Middle}: Blackbody temperature. \textit{Bottom}: Blackbody radius. Gray shaded areas represent 1$\sigma$ uncertainty.}
    \label{fig:photosphere}
    \end{figure}

\begin{figure*}[ht]
    \centering
    \includegraphics[width=\textwidth]{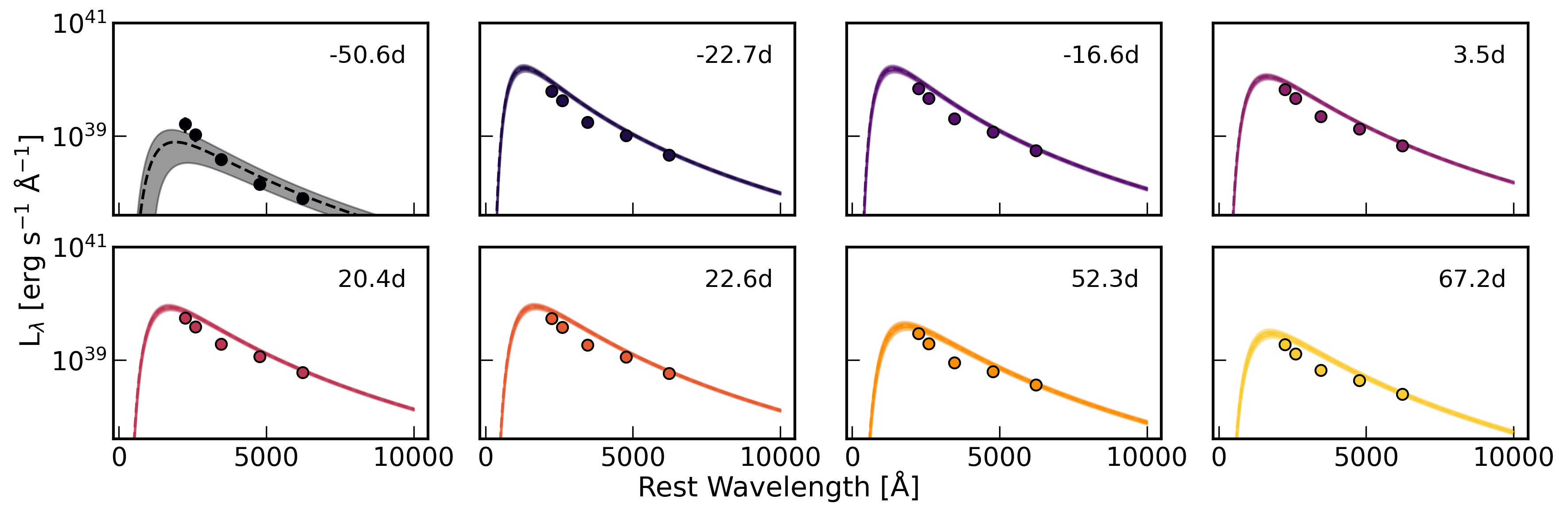}
    \caption{Blackbody fits to the SEDs.
    We show the epochs at which all five filters are available within a window of $\pm$ 1 day. 
    The top left subpanel is instead obtained with interpolated points. }
    \label{fig:sed_bb}
    \end{figure*}

    Here we summarize the workflow of the \texttt{extrabol} code \citep{thornton2024extrabol}.
    Once a set of light curves is given, the first step is to smoothly interpolate each of them using Gaussian process regression techniques.
    We excluded the $o$, and $c$ bands to avoid redundancy, as their bandpasses significantly overlap with those of other filters.
    They use a 2-dimensional kernel to interpolate the curves both in the time and wavelength domains.
    The interpolation is self-regulated and, if it results in large errors, for example when the data are very sparse, it uses template libraries to reduce the uncertainties, although the farthest from the data points, the highest the errors. 
    For this reason, the absence of UV data at times earlier than the first point produces systematically higher errors, until the fit explodes. 
    This is why we executed the fit starting from $-40$ days before $t_p$ instead of utilizing the entire time range. 
    Once the interpolation is done, magnitudes are converted into fluxes at every epoch, building an SED (Fig.~\ref{fig:sed_bb}).
    The bolometric luminosity is then obtained by integrating the SED, while temperature and radius are obtained by fitting a blackbody model to it using Monte Carlo sampling (Fig.~\ref{fig:photosphere}). 
    The default priors are $T_0 = 9000$ K and $R_0 = 1 \times 10^{15}$ cm.
    Temperature is constrained to positive values below $T_{\max} = 40000$ K, while radius can assume any positive values. See \cite{thornton2024extrabol} for more details.

\section{Host modeling with \texttt{prospector}}\label{sec:prospector}

The host galaxy contribution was modeled using the SED fitting code \texttt{prospector} \citep{prospector_package}, which models stellar population parameters from UVOIR photometry. 
Host archival magnitudes were obtained from the Sloan Digital Sky Surveys (SDSS) Data Release 16 (DR16) catalog \citep{2020ApJS..249....3A} for filters \textit{$ugriz$} and from the GALEX-DR5 (GR5) sources catalog \citep{2011Ap&SS.335..161B} for the NUV filter. 
In addition, the archival host photometry at UVW1 filter published by \cite{2025ATel17466....1M} was also included in the model. 
We adopted a parametric star-formation history including nebular emission, and sampling was performed with the \texttt{dynesty} nested sampler. 
The best-fit model SED was then used to estimate the host flux in the observed bands.
\texttt{prospector} provides an estimate of the total stellar mass of the galaxy as well as its age. 
We estimate the BH mass based on the empirical relation by \cite{reines2015relations} finding 
$\text{log}_{10} (M_{\text{BH}}) = 6.92 \pm0.55\,\rm dex$, where the uncertainty estimate is the systematic uncertainty of the relation, 
and $t_{age} \sim 3.39\,\rm Gyr$. 
The host flux was subtracted from the UVOT measurements in each corresponding bandpass using SNCosmo \footnote{\href{https://github.com/sncosmo/sncosmo}{https://github.com/sncosmo/sncosmo}}.

\begin{figure}[H]
    \centering
    \includegraphics[width=\columnwidth]{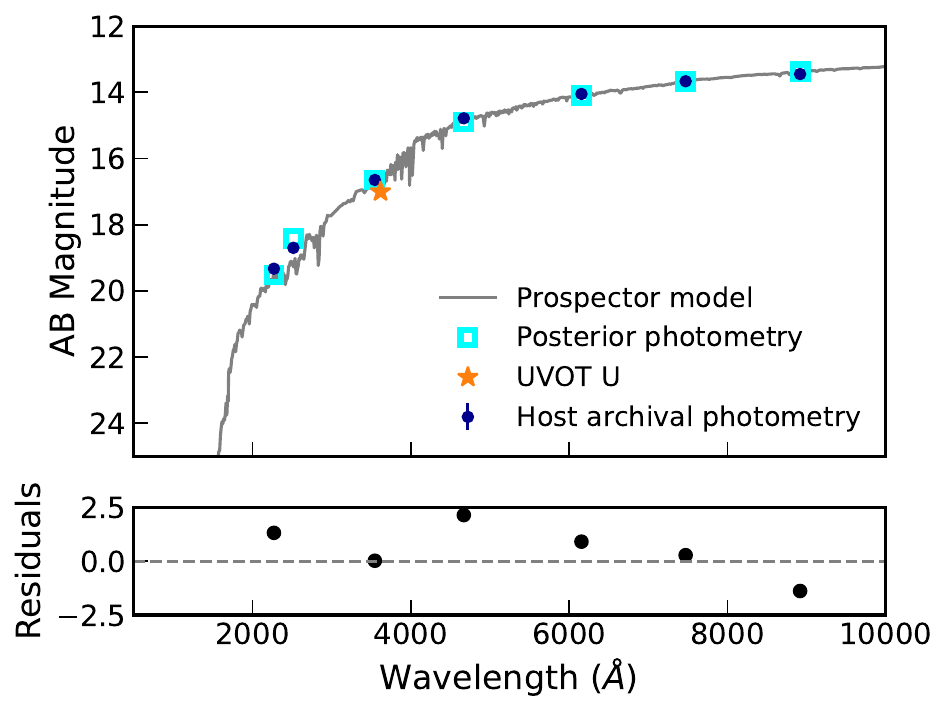}
    \caption{\textit{Top}: Host-galaxy photometry fitting with prospector and corresponding archival data. \textit{Bottom}: Residuals calculated in magnitudes.}
    \label{fig:host_fit}
\end{figure}

\begin{table}[h!]
\centering
\caption{Archival host photometry from SDSS, GALEX, and UVOT used for the construction of a synthetic galaxy model with prospector.}
\begin{tabular}{cc}
\toprule 
\toprule 
Filter    & AB magnitude       \\ 
\midrule
SDSS u    & 16.65 $\pm$ 0.01   \\
SDSS g    & 14.786 $\pm$ 0.002 \\
SDSS r    & 14.050 $\pm$ 0.002 \\
SDSS i    & 13.667 $\pm$ 0.002 \\
SDSS z    & 13.449 $\pm$ 0.003 \\
GALEX NUV & 19.3 $\pm$ 0.1     \\
UVOT W1   & 18.7 $\pm$ 0.1    \\ 
\bottomrule
\end{tabular}
\label{tab:prospector}
\end{table}

\section{Line fitting details}

\begin{figure*}[ht]
    \centering
    \includegraphics[width=\textwidth]{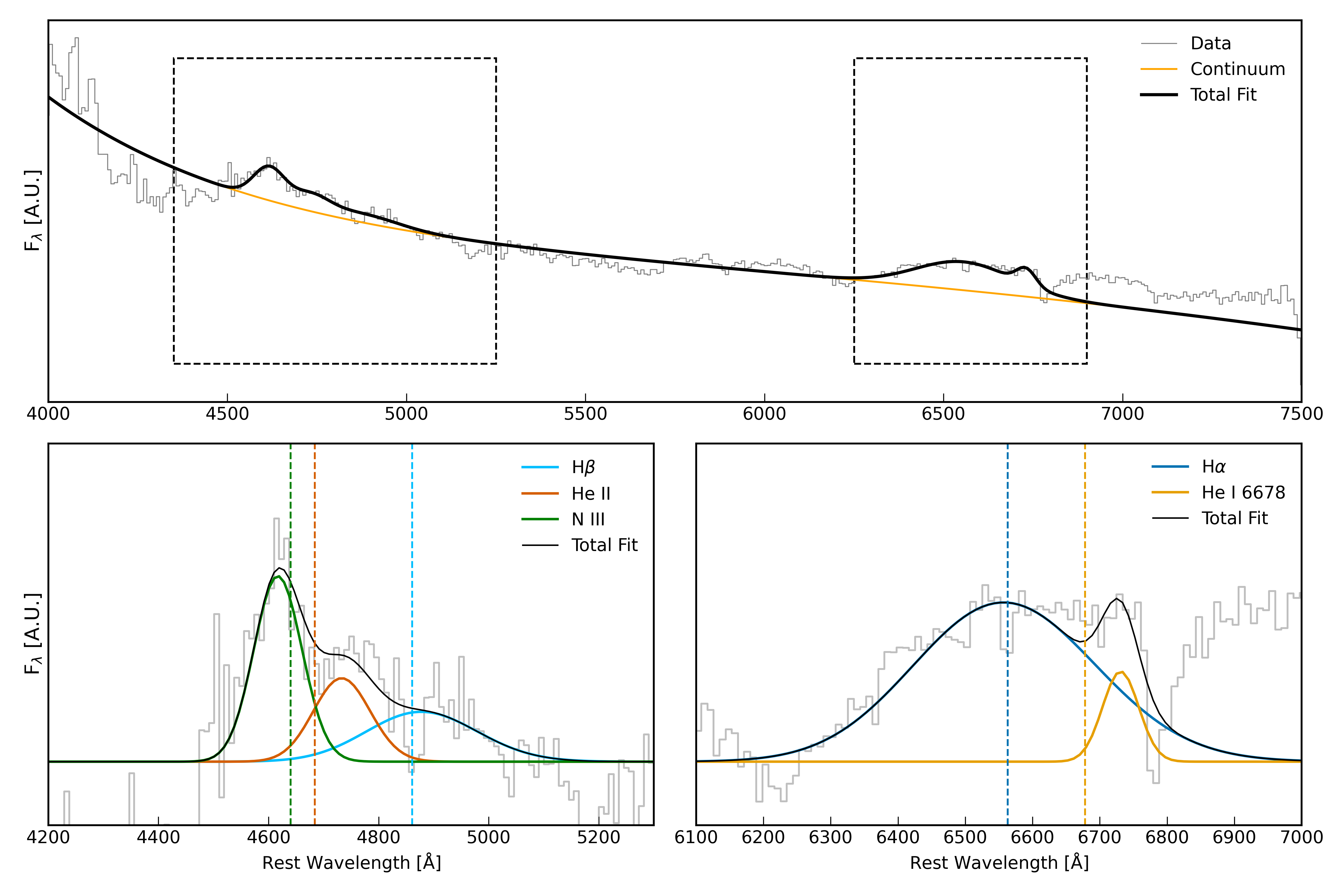}
    \caption{Visualization of multi-Gaussian fit as executed by \texttt{PyQSOFit} on our third spectrum.
    We zoomed in around the \ion{N}{iii} + \ion{He}{ii} + \ion{H$\beta$}{} complex (bottom left) and the \ion{H$\alpha$}{} + \ion{He}{i} $\lambda$6678 broad region (bottom right). }
    \label{fig:spectral_fit}
    \end{figure*}

    We fit the host+continuum subtracted spectra using a custom \texttt{python} script based on \texttt{PyQSOFit} \citep{2018ascl.soft09008G}. 
    This software allows to fit a spectrum taking as only input the spectral data, the source redshift, and a line-fitting parameter list with the prior information of each line. 
    These information include the broadness of the line, the expected centroid, and the region where to perform the fit.  
    Each individual line or blend of lines was fit with a combination of narrow or broad Gaussian features that can be specified by the user (as shown in Fig.~\ref{fig:spectral_fit}). 
    Markov-Chain Monte-Carlo routines are used to measure the uncertainties of the fitting results. 
    From the best-fit model, we derived the FWHM, estimate the line luminosity by integrating over the fit region, and computed the wavelength offset as the difference between the Gaussian peak and the rest-frame line centroid. 
    
\begin{figure}[H]
    \centering
    \includegraphics[width=\columnwidth]{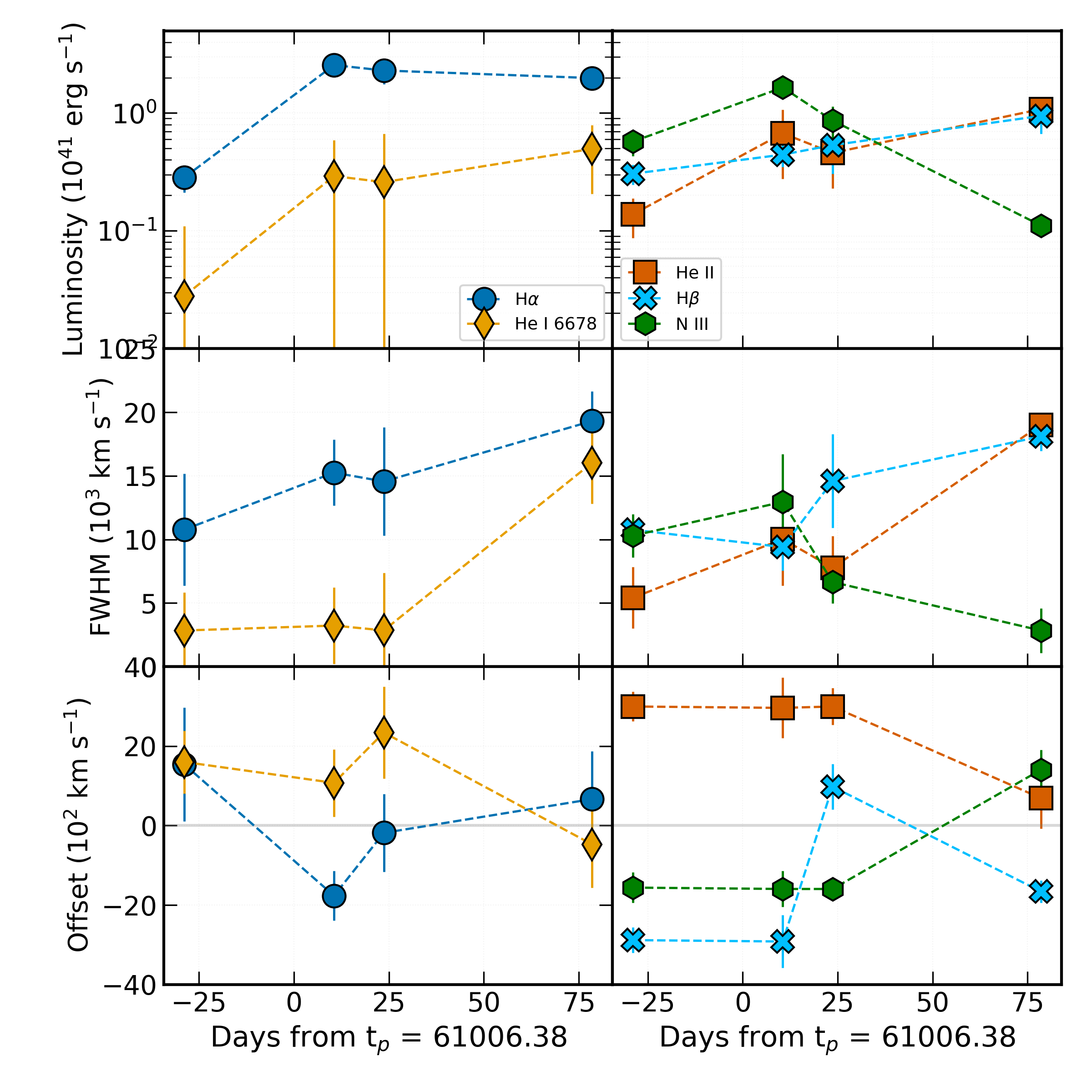}
    \caption{Luminosity, FWHM, and velocity offset of the \ion{H$\alpha$}{} + \ion{He}{i} $\lambda$6678 group and the \ion{N}{iii} + \ion{He}{ii} + \ion{H$\beta$}{} complex. 
    Note that a negative offset corresponds to a blueshift of the line. }
    \label{fig:lines}
    \end{figure}

    Results of line fitting are shown in Fig.~\ref{fig:lines}. 
    Despite the fit manages to individuate the contribution from \ion{N}{iii}, \ion{He}{ii}, and \ion{H$\beta$}{} at every step, we note that the line is heavily blended, hindering a perfect separation of the two. 
    This could, for instance, be reflected in the high redshift (and, conversely, high blueshift) of the \ion{He}{ii} (\ion{N}{iii}) line during the first epochs.
    Similarly, the \ion{H$\alpha$}{} and \ion{He}{i} $\lambda$6678 features are strongly degenerate, although being more separated. 
    For these reasons, these results serve mostly as a general indication of the evolution of the spectral lines, more than a precise measurement.
    The \ion{He}{i} $\lambda$5876 line is only marginally detected and appears a strong feature (FWHM$\sim 20000\,\rm km\,s^{-1}$, $L_{\rm He\, I} \sim 3\times 10^{40}\,\rm erg \, s^{-1})$ only in our first spectrum. 
    Notably, the \ion{He}{ii} / H$\alpha$ luminosity ratio assumes the following values: $0.49\pm0.22$, $0.26\pm0.15$, $0.20\pm0.11$, and $0.54\pm0.03$. 
    As commented in the main text, this evolution inversely mirrors the evolution of the photometric radius.

\clearpage

\section{Extra tables and figures}

\begin{figure}[h!]
    \centering
    \includegraphics[width=\columnwidth]{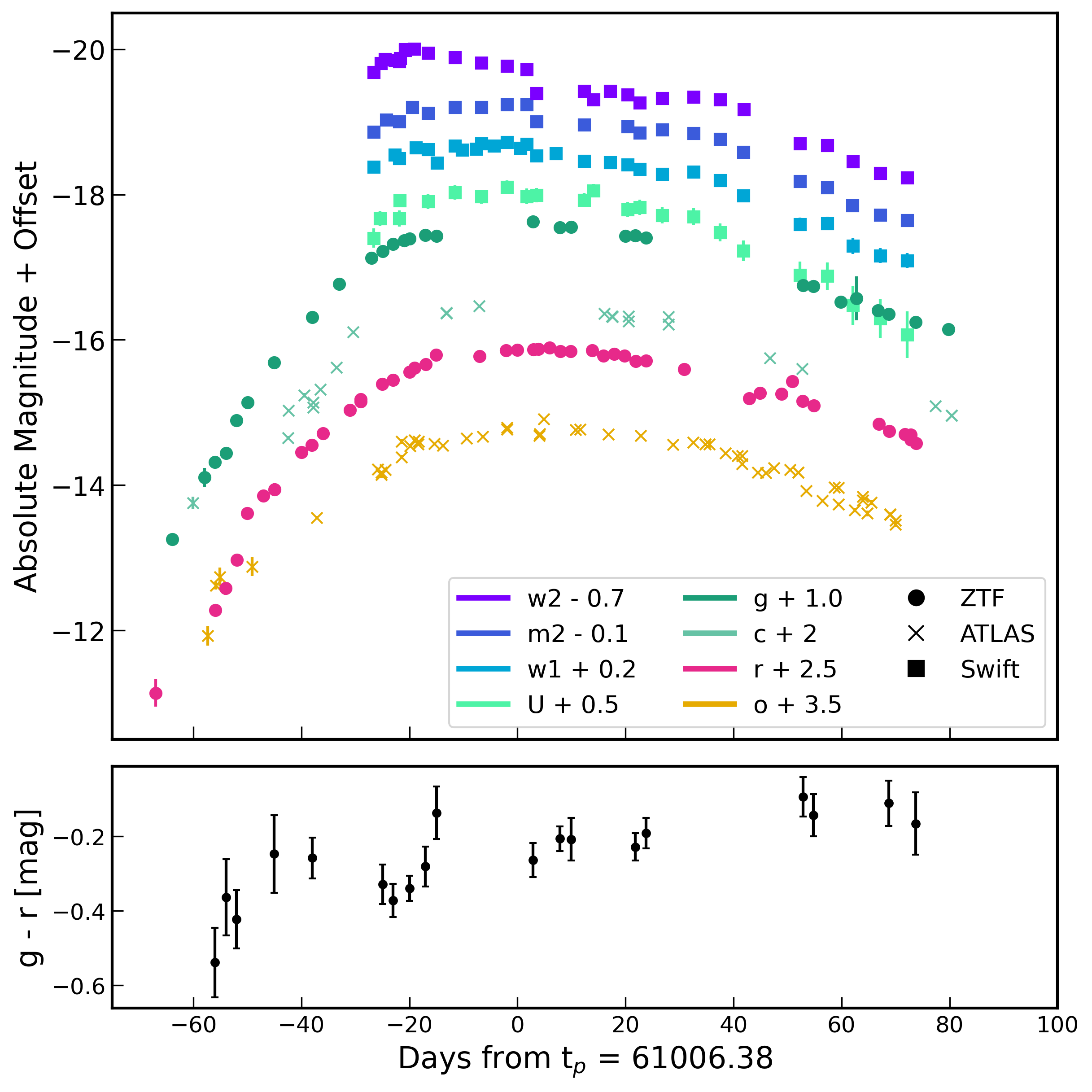}
    \caption{\textit{Top}: Light curves of TDE~2025aarm during the first $\sim$ 150 days after discovery. 
    Filters are offset for visual clarity. 
    \textit{Bottom}: $g-r$ color evolution.}
    \label{fig:lightcurves}
\end{figure}

 \begin{table*}
    \centering
    \caption{Input and output parameters for the \texttt{tde} model from \texttt{MOSFiT}.}
    \begin{tabular}{l|ccc}
    \toprule
    \toprule
    Parameter & Prior range & Prior type & Posterior \\
    \midrule
    $M_\star$ [$M_\odot$]            & [0.1 -- 10]                      & Uniform       &  $0.155_{\small -0.0309}^{\small +0.0297}$        \\ 
    $M_\text{BH} ~[10^6M_\odot]$ & [$5\times10^{-1}$ -- $1\times10^2$]  & Log-uniform   &  $18.74^{\small +3.25}_{\small -2.22}$ \\ 
    $t_{\rm dis}\,{\rm [day]}$   & [-30, 0]                             & Uniform       &  $-18.92_{\small -2.6}^{\small +2.01}$ \\ 
    $T_{\rm viscous}$ [day]      & [$10^{-2}$ -- $10^2$]                & Log-uniform   &  $17.38_{\small -8.37}^{\small +6.1}$  \\ 
    b                            & [0.5 -- 1.5]                         & Uniform       &  $0.54_{\small -0.029}^{\small +0.072}$ \\
    $\epsilon$                   & [0.03 -- 0.3]                        & Log-uniform   &  $0.03575_{\small -0.00441 }^{\small +0.00773}$  \\
    $l$                          & [1 -- 2]                             & Uniform       &  $1.847 _{\small -0.172 }^{\small +0.097}$\\
    $R_{ph0}$                    & [$10^{-4}$ -- $10^4$]                & Log-uniform   &  $156.6^{\small +110}_{\small -72.2}$  \\ 
    $\sigma$                     & [0.01, 10]                           & Log-uniform   &  $0.2241_{\small -0.097 }^{\small +0.0178}$ \\
    \bottomrule
    \end{tabular}
    \tablefoot{The fit was executed giving as input redshift, Galactic extinction, line-of-sight hydrogen column density, luminosity distance, number of walkers (100) and number of iterations (50000). 
    We used the same nomenclature as in \protect\citep{mockler2019weighing}, where: $M_\star$ is the stellar mass; $M_\text{BH}$ is the BH mass; $t_{\rm dis}$ is the time since first detection; $T_{\rm viscous}$ is the viscous delay time; $b$ is the scaled impact parameter; $\epsilon$ is the efficiency; $l$ is the photosphere power-law exponent; $R_{ph0}$ is the photosphere power-law constant; $\sigma$ is the variance. 
    }
    \label{table:mosfit}
\end{table*}

 \begin{table*}[!]
    \centering
    \caption{Log of spectral observations.}
    \label{tab:spectra}
    \begin{tabular}{l|ccccccc} 
    \toprule 
    \toprule  
    Date       & Phase & $\lambda_\text{min}$ & $\lambda_{\text{max}}$ & Exposure time & $\bar{\Delta  \lambda}$ & Instrument & Telescope \\ 
    MJD        & (day)   & ($\AA$)       & ($\AA$)                  & (s)        &   [$\AA$]         &             \\ 
    \midrule 
    60977.43 & -28.78 & 3200 & 10000 & 2700 & 2  & FLOYD & LCO \\ 
    61016.93 & 10.72 & 4000 & 8000  & 1600 & 10 & SPRAT & LT \\
    61030.04 & 23.82 & 4000 & 8000  & 1600 & 10 & SPRAT & LT \\
    61084.90 & 78.68 & 4000 & 8000  & 1600 & 10 & SPRAT & LT \\
    \bottomrule 
    \end{tabular}     
    \tablefoot{Phases are expressed with respect to the $g$-band maximum $t_{\rm p}$ in the observed frame.}
\end{table*}

\begin{figure}[h!]
    \centering
    \includegraphics[width=\columnwidth]{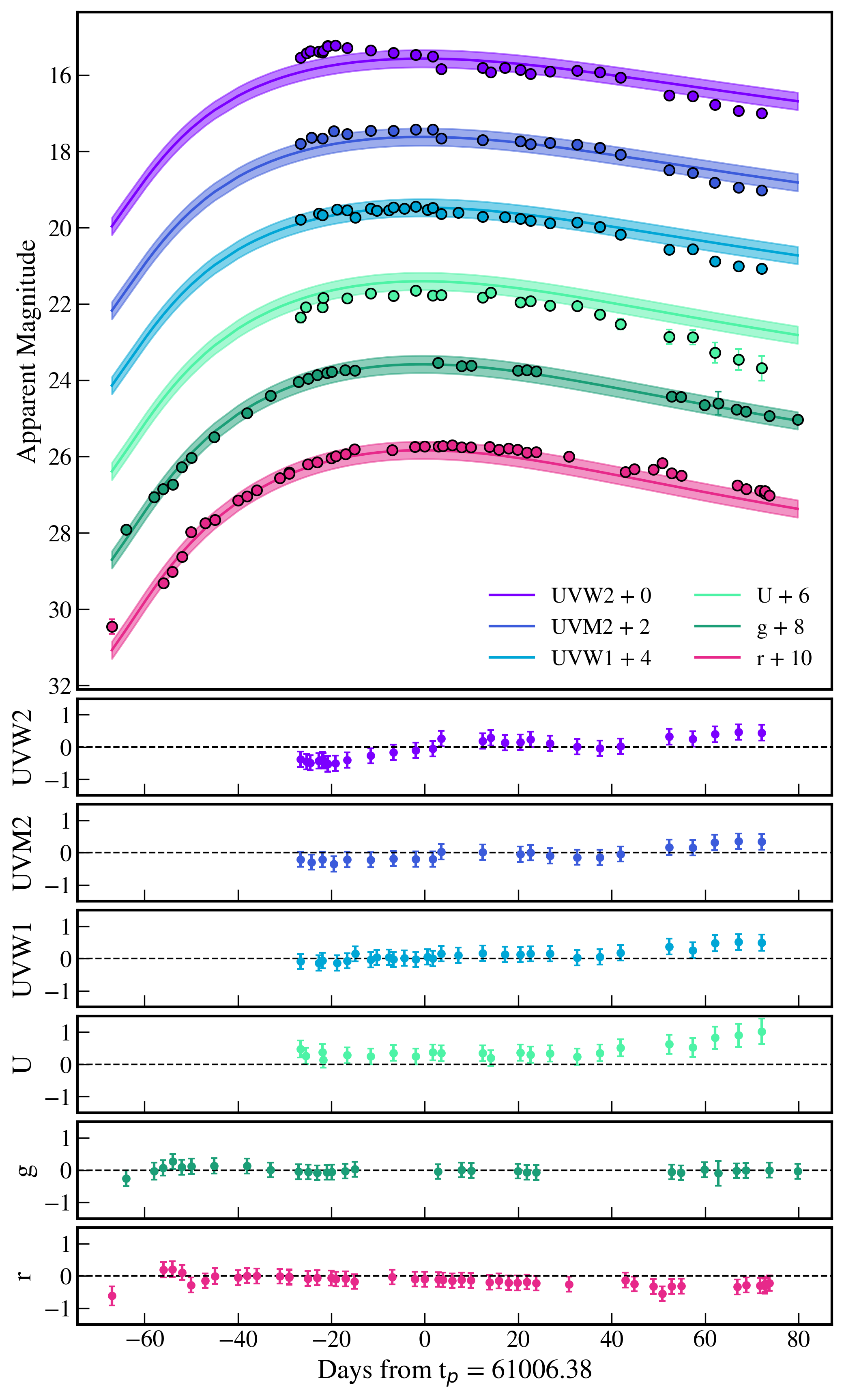}
    \caption{Fallback modeling of the optical light curves of TDE~2025aarm using \texttt{MOSFiT}. \textit{Top panel}: Points identify our data, while the lines are the fit. Light curves are offset for visual clarity. \textit{Bottom panels}: Residuals of the fit in mag units.}
    \label{fig:mosfit}
\end{figure}

 \begin{table*}[!]
    \centering
    \caption{Photometric data of TDE~2025aarm (extract).}
    \label{tab:uvot}
    \begin{tabular}{l|ccccc} 
    \toprule 
    \toprule  
    Date & Phase & Mag & Err & Band  & Telescope \\ 
    MJD  & day   & AB  & AB  &       & \\ 
    \midrule 
    60939.42 & -66.97 & 20.46 & 0.01 & r & ZTF \\
    60942.46 & -63.92 & 19.92 & 0.02 & g & ZTF \\
    60946.28 & -60.10 & 18.39 & 0.00 & c & ATLAS \\
    60948.44 & -57.94 & 19.07 & 0.02 & g & ZTF \\
    60949.07 & -57.31 & 18.65 & 0.01 & o & ATLAS \\
    60950.34 & -56.05 & 18.86 & 0.02 & g & ZTF \\
    60950.42 & -55.97 & 19.32 & 0.01 & r & ZTF \\
    60950.55 & -55.83 & 17.96 & 0.01 & o & ATLAS \\
    60951.28 & -55.10 & 17.84 & 0.01 & o & ATLAS \\
    60952.40 & -53.98 & 19.02 & 0.01 & r & ZTF \\
    :        &  :     & :     &  :   & : & : \\ 
    :        &  :     & :     &  :   & : & : \\ 
    \bottomrule 
    \end{tabular}     
    \tablefoot{Phases are expressed with respect to the $g$-band maximum $t_{\rm p}$ in the observed frame. The full table is available at the CDS.}
\end{table*}

\end{appendix}
\end{document}